\documentclass{kapproc} 

\usepackage{procps}
\usepackage{amssymb} 

\usepackage[dvips]{graphicx}

\upperandlowercase

\kluwerbib 

\normallatexbib 

\begin{document}



\articletitle{Stellar explosions: from supernovae to gamma-ray bursts}

\chaptitlerunninghead{SNe and GRBs}















\author{Konstantin Postnov\altaffilmark{1,2}}

\altaffiltext{1}{Sternberg Astronomical Institute, Moscow, Russia}
\altaffiltext{2}{University of Oulu, Finland}
\email{pk@sai.msu.su}




\begin{abstract}
Current understanding of core collapse and
thermonuclear supernovae is reviewed. 
Recent 
progress in unveiling the nature of cosmic gamma-ray bursts 
(GRB) is discussed, with the focus on the apparent link of several GRBs with 
an energetic subclass of stellar explosions, type Ib/c core-collapse
supernovae. 
This relation provides the strong case that the GRB phenomenon 
is connected with the final stages of massive star evolution 
and possibly with the formation of neutron stars and black holes.

\end{abstract}


\section*{Introduction}
Seventy years ago W. Baade and F. Zwicky (1934) \cite{Baade&Zwicky1934} 
were the first to point out
that one of the brightest astronomical phenomenon, supernovae stars (SNe), 
can be
due to explosions of massive stars at the end of their evolution. The
formation of a dense neutron core (neutron star) results in a sudden energy
release of order of the gravitational binding energy of the neutron star
which amounts to $E_g\sim -GM_{NS}^2/R_{NS}\approx 10^{53}$ ergs for the
canonical values of the NS mass $M_{NS}\approx 1 M_\odot$ and radius
$R_{NS}\approx 10$ km. It was soon recognized by 
Gamow and Schoenberg (1941) \cite{Gamow&Schoenberg1941} that 
most of this energy comes into neutrino emission. Twenty years after, Hoyle
and Fowler (1960) \cite{Hoyle&Fowler1960} 
showed that energy released in 
type Ia supernovae are connected to thermonuclear
burning of a degenerate stellar core. Here the available energy is $\sim 0.007 Mc^2
\approx 2\times 10^{51}$ ergs for 
the Chandrasekhar mass of the white dwarf.

In the end of 1960s cosmic gamma-ray bursts (GRBs) were discovered
by gamma-ray satellites \cite{Klebesadel&al1973, Masetz&al1974}. 
Largely due to the inability of precise localization of
the GRB position on the sky
using gamma-ray detectors only, the
origin of GRBs was as enigmatic as that of SNe before Baade and
Zwicky's suggestion until late 1990s, when the first successful 
localizations of GRBs using their afterglow emission were made by BeppoSAX
satellite in X-rays \cite{Costa&al1997}. The detection of X-ray afterglows
several hours after GRB allowed 
dedicated follow-up observations of the GRB error boxes 
to be carried out using powerful optical 
(e.g. \cite{jvP&al1997}) and radio telescopes \cite{Frail&al1997}, 
in which rapidly decaying afterglow emissions were also detected. 
Quite soon after that, in the spring 1998, a bright peculiar 
nearby supernova 1998bw was found within the error box 
of GRB 980425 \cite{Galama&al1998, Kulkarni&al1998}, suggesting the link 
between GRBs and SNe. 
Presently we have several unequivocal associations of GRBs
with very energetic type Ibc supernovae called "hypernovae" 
(see below, Section \ref{sec:sn-grb}).     

Being connected to the evolution
of stars, SN studies overlap with practically all fields of
the modern astronomy, from physics of tiny interstellar medium
(e.g. \cite{Lozinskaya}) to the formation 
of superdense neutron stars \cite{Lattimer&Prakash2004}. As was
also pioneered by Baade and Zwicky, they are sources of astrophysical shocks
in which cosmic ray particles are accelerated \cite{Blandford&Eichler1987}. 

Here we focus on some recent highlights in both
core collapse and thermonuclear supernova studies, which became possible
mainly due to increasingly accurate radiation hydrodynamic calculations 
with a detailed treatment of neutrino processes. We also briefly
describe recent success of asymmetric SN simulations (2D magneto-rotational
collapse).
Next we focus on recently
established link between GRB explosions and energetic type Ibc supernovae
(hypernovae) and discuss recent ideas on
the GRB progenitors. We 
hypothesize that different core collapse outcomes may lead
to the formation of different classes of GRBs.

\section{Core collapse supernovae}

An extensive discussion of basic physics of core collapse supernovae can be
found in \cite{Bethe1990}; the evolution of massive stars, 
core collapse, formation of stellar remnants 
and supernova nucleosynthesis 
are reviewed in \cite{Woosley&al2002}; a recent concise discussion of 
problems and prospects for core collapse supernovae can be found in  
\cite{Mezzacappa2004}. 

In the end of thermonuclear evolution, the core of a massive star can lose 
mechanical stability for various reasons. In the stellar mass range 
$8M_\odot<M<20 M_\odot$ 
a partially degenerate core with mass close to the 
Chandrasekhar limit $M_{core}\sim M_{Ch}$
and high density ($\rho\sim 10^9-10^{10}$ g/cm$^3$) appears. Under these
physical conditions, the chemical potential of degenerate electrons becomes so
high that neutronisation reactions 
$ e^-+(A,Z)\to (A,Z-1) +\nu_e $
become effective even at zero temperature.
At  densities $>10^6$ g/cm$^3$ degenerate electrons
becomes
relativistic so the adiabatic index of matter  
$\gamma=d\log P /d\log \rho \to 4/3$, the critical value for loss of
mechanical stability. 
Neutronisation of matter means its deleptonisation (decrease in
the lepton number $Y_e=N_e/N_b$), so the pressure at some moment increases
slower than $\rho^{4/3}$ though $\gamma$ is formally above $4/3$ calculated
at constant $Y_e$, and a catastrophic collapse begins. 
Temperatures are higher at larger stellar masses  $M>20 M_\odot$ so the
collapse is initiated by  photodissociation of nuclei (here $\gamma$ really
becomes $<4/3$). For most massive stars with $M>60 M_\odot$ pair creation 
makes $\gamma<4/3$.

The collapse occurs on the dynamical (free-fall) time scale $t_{ff}\sim
1/\sqrt{G\rho}\sim $ a fraction of second.
Adiabaticity holds so the entropy per baryon $s=S/k_B\approx const\sim 1$
and even can increase due to non-equilibrium beta-processes.
Low specific entropy (compared to H-burning phase at the main sequence, 
$s\approx 10-15$) 
prevents dissociation of nuclei until they
"touch" each other at densities of the order of the nuclear density,
$\rho_n\sim 2\times 10^{14}$ g/cm$^3$.
The collapse stops (if the core mass is below some $M_{max}$), 
and bounce of the shock occurs at $\sim 50$ km from the center.

The bounce shock heats up deleptonized matter and rapidly 
spends most of its kinetic energy to
destroy nuclei and produce plenty of free nucleons (n, p).  
Modified URCA-processes \cite{Dicus1972} 
becomes important: $e^-+p \to n+\nu_e$, $e^++n \to p+\bar \nu_e$
and pair-neutrino annihilation takes place:
$e^++e^-\to \nu_e+\bar \nu_e$. 
At the typical collapse temperatures 
$T\sim 10$ MeV a lot of $\nu$'s is produced \cite{Zeldovich&Gusseinov1965}
However, at densities $\rho\sim 10^{12}$ g/cm$^3$ the mean free path of 10
MeV neutrinos is by 5-6 orders of magnitude smaller than the size of the
proto neutron star ($R\sim 50$ km) so the opaque "neutrinosphere" forms.
Most of the core collapse neutrinos diffuse out of the neutrinosphere on a
time scale $\sim 10$ seconds. First calculations of $\nu$ spectra in core
collapse SN were performed by D.K. Nadyozhin
\cite{Nadyozhin1978, Nadyozhin&Otroshenko1980}
We should note that subsequent detailed calculations (e.g.
\cite{Messer&al1998} and references therein) did not change much
these spectra. Thus, the modest $\sim 10\%$ fraction of the
total neutrino energy released in the core collapse ($\sim
10^{53}$ ergs) would be sufficient to unbind the overlying
stellar envelope and produce the phenomenon of type II supernova
explosion.

\subsection{Neutrino-driven explosions}

Thermal SN explosion mechanism was proposed by 
Colgate and White \cite{Colgate&White1966}. In this picture, 
part of the neutrino flux liberated in the core collapse is deposited
to the stellar mantle to make it unbound ($\sim 10^{51}$ ergs is needed).
Specific mechanisms include neutrino-driven fluid instabilities,
for example convection both 
above neutrinosphere and inside the proto-NS.
Neutrino-driven convection, however, may not be as important as thought before,  
as follows from recent detailed 2D studies of convection  
\cite{Buras&al2003}. Instead, other fluid instabilities 
such as newly found double-diffusive instability (the so-called 
"lepto-entropy fingers") \cite{Bruenn&al2004}, may effectively 
increase neutrino luminosity to help successful explosion. 
Note here that process $\nu_e+\bar \nu_e\to e^++e^-\to \gamma+\gamma$
may also be important above $\nu$-sphere 
\cite{Berezinsky&Prilutsky1987}. 
This process was proposed as the energy source for GRB fireballs. 

It is now realized that even detailed 1D-calculations of the core collapse
supernovae fail to produce explosion (see e.g. reviews by 
\cite{Libendoerfer2004, Mezzacappa2004} and references therein).
The main reasons
are that the bounce shock rapidly stalls (turns into a pure accretion front) 
over neutrinosphere at $R_s\sim
200$ km because of nuclear dissociations, and net neutrino heating of the
freshly accreted material immediately after the front is insufficient. 
Burrows \cite{Burrows2004} 
notes that 1D-models are about to produces a successful explosion, 
only 25\% to 50\% increase in 
the energy deposition rate inside the "gain region" is required.
The stalled shock is not revived because the fresh fall-back matter
is advected more rapidly than heated up by neutrinos.  
An additional heating by the viscous dissipation 
in differentially rotating models is considered
in  \cite{Thompson&al2004}. Differential rotation of collapsing
core seems quite plausible (see evolutionary calculations of rotating stars  
\cite{Heger&al2000, Heger&al2004}). 
The energy stored in shear can amount to $\sim
10^{52}$ ergs for reasonably short 6 ms spin periods of 
proto neutron stars. 
The magneto-rotational instability (MRI) which 
arises in differentially rotating magnetized fluids with negative 
angular velocity gradients $d\Omega/dr<0$ 
\cite{Velikhov1959, Chandrasekhar1960}. 
MRI is thought to be responsible for 
turbulence in accretion disks and may
operate in stars as well. This viscosity was found
\cite{Thompson&al2004} to be comparable with neutrino-driven convective
viscosity and much larger than the neutrino viscosity. The account for
additional energy dissipation due to the MRI viscosity in differentially
rotating stellar cores allowed \cite{Thompson&al2004} 
to obtain a successful explosion in their 1D-calculations.

Transition from 1D to multi-dimensional hydrodynamic calculations
has enabled a more accurate treatment of fluid instabilities. 
Presently, it is recognized 
that even with detailed account of Boltzmann neutrino transport,
additional physical properties of the collapsing core, e.g. rotation, 
need to be coupled with multi-dimensional
calculations to produce successful neutrino-driven 
explosions (see e.g. \cite{Janka&al2004a, Mezzacappa2004}).  

\subsection{Asymmetric explosions}

There are increasing observational evidence that supernova explosions
are generically asymmetric. The indications are as follows. 

\paragraph{Spectroscopic observations}

The asphericity of the explosion 
is directly inferred from spectropolarimetric observations
\cite{Leonard&Filippenko2004}. The smaller the hydrogen mass of the 
presupernova envelope, the higher the degree of linear polarization
is observed. In the canonical type II-plateau supernovae with heavy 
hydrogen envelopes (SN1999em, 2003gd etc.) 
polarization is small (less than 0.5\%), indicating
low degree of asphericity. In contrast, in the stripped-envelope progenitors
giving rise to type IIb (SN1993J), type IIn (SN1998S), type Ic (SN2003gf)  
or peculiar Ic (SN2002ap, maybe hypernova, see below) supernovae,
th polarization degree  is large, of order of 1-2\%. 
The degree of polarization tends to increase with time
(i.e. as more and more deeper layers of the ejecta become transparent).
This suggests 
the relation of asymmetric supernova explosion to pulsar kicks and, in
extreme cases, to GRB explosions (see below).   

\paragraph{Pulsar kick velocities}

Based on measurements of proper motions of $\sim 100$ radio pulsars,  
it is well established now 
\cite{Lyne&Lorimer1994, Arzoumanian&al2002} that 
young neutron stars demonstrate space velocities up to several 100 km/s which 
cannot be explained without introducing additional natal kicks.  

\paragraph{Shape of young galactic SNR}

The asphericity  of SNRs 1987a, Cas A, N132D
etc. is directly seen in multiwavelength observations
\cite{Park&al2002, Hwang&al2004}. Recent 
\textit{Chandra} X-ray observations of galactic SNR
W49B \cite{W49B} suggest this SNR may be the
remnant of an asymmetric hypernova explosion in our Galaxy.   

\paragraph{Strong mixing of Ni-56}

Strong mixing of $Ni-56$ in the outer layers of the ejecta 
is required to explain the observed SN light curves (especially SN1987a,
\cite{Mitchell&al2001}). 

\vskip\baselineskip
Different mechanisms for the SN asymmetry have been proposed. 

\paragraph{Hydrodynamic mechanism} 

Large-scale neutrino-driven overturn
between the shock and neutrinosphere  
can cause the SN anisotropy explosion \cite{Burrows&al1995}.
Pulsar kicks up to 500 km/s from neutrino convection
were obtained in recent 2D hydrodynamic calculations
with Boltzmann neutrino transport 
\cite{Buras&al2003}, while in 3D-calculations (but with 
less rigorous treatment of neutrino transport) \cite{Fryer2004} 
high initial velocities of proto-NS were found 
to be damped by neutrino emission generally 
to less than 200 km/s. However,  
the postbounce stellar core flow is found (both in 2D and 3D) 
to be subjected to generic non-spherical perturbations
of the accretion shock leading to the development of the so-called
"stationary accretion shock instability" \cite{Blondin&al2003}. 
This large-scale ($l=1,2$ modes) 
fluid instability was also found in multi-dimensional 
simulations in Ref. \cite{Janka&al2004b}, and may be responsible for observed
bimodal pulsar velocity distribution \cite{Arzoumanian&al2002}. 
It is important that 
the developed accretion-shock instability can result in bipolar
explosions even in the absence of rotation. 

\paragraph{Magnetorotational mechanism} 

Rotation of the pre-supernova core and magnetic field increase during the
collapse, and was proposed in the beginning of 1970s \cite{BK1970} as an
alternative mechanism for core-collapse SNe. The magnetorotational mechanism
by its nature is asymmetric and can launch antiparallel jets during the
explosion. First simulations were made in Ref. \cite{LeBlanc&Wilson1970,
Symbalisty1984}. Results of recent 2D MHD calculations
\cite{Ardeljan&al2004a, Ardeljan&al2004b} are very encouraging: strong
differential rotation in the presence of (maybe initially weak) magnetic
field was shown to increase the magnetic pressure and form a MHD shock. As a
result, rotational energy of neutron star is converted to the energy of the
radial expansion of the envelope. The exponential growth of magnetic field
in differentially rotating collapsing stellar cores due to MRI
\cite{Akiyama&al2003} appears to help the magnetorotational explosion as
well. The magnetorotational explosion was found in
\cite{Ardeljan&al2004b} to be
essentially divided into three stages: linear growth of the toroidal
component of the magnetic field due to twisting of the magnetic filed lines, 
the exponential growth of both toroidal and poloidal field components due
to development of MHD instabilities, and the formation of a MHD shock 
leading to the explosion. These 2D MHD simulations \cite{Ardeljan&al2004b}
have shown that the obtained energy of the magnetorotational explosion 
$\sim 0.6\times 10^{51}$ ergs is sufficient to explain type II and type Ib
core-collapse supernovae.     

\paragraph{Formation and subsequent collision of a close binary NS+NS system inside
the stellar interiors} 

V.S. Imshennik \cite{Imshennik1992} suggested the following scenario: 
a rapidly rotating core collapse results in the 
core fusion due to rotational instability
(the original idea goes back to von Weizsaecker \cite{vonWeizsaecker1947}),
then the binary NS system  
coalesces due to gravitational radiation losses.
In this scenario, the lighter NS (with larger radius)
first fills its Roche lobe and start losing mass as the orbit shrinks due to
gravitational radiation. The NS gets unstable when
$M<M_{min}\approx 0.1M_\odot$ 
providing an energy release of $\sim 10^{50}$ erg
in neutrinos 
\cite{Blinnikov&al1984, Colpi&al1989, Colpi&al1991}. 
This mechanism has been further elaborated in Ref.
\cite{Imshennik&Popov1994, Aksenov&al1997}. It was suggested for SN1987a explosion 
\cite{Imshennik&Ryazhskaya2004} to explain double LSD neutrino signal
separated by $\sim 5$ hours  
observed from SN1987a. A point of concern here may be a rapidly rotating
pre-supernova core required for this mechanism to operate. 
Recent calculations by the Geneva group of stellar evolution with rotation 
but without magnetic field \cite{Hirschi&al2004} indicate that 
there are enough angular momentum in the pre-supernova core, while 
calculations with (even approximate) account for magnetic field  
\cite{Heger&al2004} show a severe (30-50 times) reducing of the final rotation
of the collapsing iron core in massive stars.
The core fragmentation during collapse with rotation was also not 
found in 3D calculations in Ref. \cite{Fryer&Warren2004}. 
 
\paragraph{Non-standard neutrino physics} 

To explain high pulsar kicks, neutrino
asymmetries in high magnetic fields pertinent to young pulsars 
have been invoked \cite{Chugaj1984, BK1996, Lai&al2001}.  
The neutrino asymmetry is a natural consequence of 
the asymmetry of basic weak interactions in the presence of 
a strong magnetic field. Neutrino oscillations requiring a sterile 
neutrino with mass 2-20 keV and a small mixing with active neutrinos
was proposed in Ref. \cite{Kusenko2004}. However, a critical analysis
in Ref. \cite{Janka&al2004b} suggests that 
neither rapid rotation nor strong magnetic field of the young neutron star
can be unequivocally inferred from observations of core-collapse 
SNR 1987A and Cas A, so the solution to the 
problem of high pulsar kicks 
probably should be looked for within the frame of
the conventional hydrodynamic SN explosion mechanism (see above). 
 
\section{Thermonuclear supernovae}

Thermonuclear supernovae constitute a separate very important class of
stellar explosions. In contrast to core-collapse SNe they demonstrate very
similar light curves and which allowed their using as "standard candles" in 
modern cosmology (e.g. \cite{Riess&al2004} and references therein). 
SN Ia are due to thermonuclear burning of a (C+O) white dwarf with $M\sim M_{Ch}$
\cite{Hoyle&Fowler1960}. A recent deep review of the SN 1a 
explosion models can be found in Ref. \cite{Hillebrandt&Niemeyer2000}.
In the modern picture (e.g. 
\cite{Woosley&al2004}), 
thermal instabilities in degenerate matter with
$\rho\sim 2-9\times 10^9$ g/cm$^3$, $T\sim 7\times 10^8$ K,
after $\sim 1000$ years of core convection initiate 
flame ignition within a typical ignition radius $\sim 150-200$ km
and could result in the 
complete destruction of 
the white dwarf. 
In principle, at higher density
a neutron star could be formed due to electron-capture processes 
(via accretion-induced collapse of
a O+N+Mg core or accretion onto a O-Ne-Mg white dwarf in a binary 
system \cite{Schatzman1962, Iben&Whelan1973}), but higher densities are disfavored from evolutionary
viewpoint. 

Possible scenarios for the formation of type Ia SNe include double
degenerate white dwarf mergings or accretion onto a massive white dwarf in a
symbiotic binary (see discussion of different formation channels for SN Ia
progenitors in Refs. \cite{Yungelson&Livio1998, Ruiz-Lapuente&al2000,
Yungelson2004}). Sub-Chandrasekhar mass models for SN1a due to an external
trigger (e.g. detonation of the accreted He layer) may result in subluminous
SN Ia (like 1991bg). A SN 1.5 model in which the degenerate core explodes at
the late asymptotic giant phase of the evolution of an intermediate mass
star was suggested in Ref. \cite{Iben&Renzini1983} SN 2002ic and SN 1997cy
in whose spectra hydrogen absorption lines were discovered probably belong
to this class \cite{Chugai&Yungelson2004}. In this model the hydrogen
envelope ejection is synchronized (within $\sim 600$ years) with the
explosion of a contracting white dwarf as its mass approaches the
Chandrasekhar limit.

After the initial thermonuclear ignition of the degenerate core
interior, a strong temperature dependence of the nuclear
reaction rates ($\propto T^{12}$ at $\sim$ MeV temperatures)
leads to the formation of very thin burning layers propagating
conductively with subsonic speeds (deflagration, i.e. a flame)
or burning due to shock compression (supersonic detonation).
Both propagation modes are linearly unstable to spatial
perturbations and presently are treated using multi-dimensional
calculations. The prompt detonation of the degenerate core as SN
1a mechanism was first studied by D. Arnett \cite{Arnett1969}.
However, the mechanism is inconsistent with observations as too
little intermediate-mass elements are born in detonation, and
was found not to operate because the core at ignition is
insufficiently isothermal \cite{Woosley1990}. Deflagration to
detonation (delayed detonation) burning regime was suggested in
Ref. \cite{Ivanova&al1974} and further elaborated in
\cite{Khokhlov1991}. Pure carbon deflagration with convective
heat transport was proposed in Ref. \cite{Nomoto&al1976}.

The main problem encountered in studies of the SN Ia explosion
mechanisms can be formulated as follows: 
the prompt detonation produces enough energy for explosion 
(thermonuclear runway energy a white dwarf 
$\sim 0.007Mc^2 \sim 1.5\times 10^{51}$ ergs) 
but gives incorrect nucleosynthesis yields (mainly iron peak elements, in   
contrast with observations showing significant 
abundance of the intermediate mass elements), the  
pure deflagration is too slow and must be accelerated.
The acceleration of the deflagration front is achieved by involving
different flame instabilities:  
Landau-Darrieus flame instability, which leads to the flame front
fractalization - wrinkles and folds, so the surface area of the flame
effectively increases
(see analytical treatment in 
\cite{Blinnikov&Sasorov1996} and numerical simulations 
in \cite{Roepke&al2004});
(2) Rayleigh-Taylor (RT) instability resulting from the buoyancy of 
hot, burned material in the dense, unburned surroundings.
The RT instability generically develops after the initial ignition in 
degenerate cores turbulent deflagration and leads to the appearance of 
the hot bubbles ("mushrooms") floating upwards while spikes of cold fluid
falls down. Secondary hydrodynamic instabilities due to shear along the bubble
surface rapidly leads to the turbulence. The key 
role of the RT instability in the presently concurrent 
turbulent deflagration and
delayed-detonation models for SN Ia explosions is 
fully confirmed in 3D numerical simulations
\cite{Reinecke&al2002, Gamezo&al2003}.  

Recent progress in the physics of the 
SN Ia explosions has been done mainly due to
multi-dimensional numerical simulations of the flame propagation in the
degenerate star. For example, multidimensional Chandrasekhar mass
deflagration simulations
\cite{Reinecke&al2002}
indicate acceleration of the turbulent combustion front up to
30\% of the speed of sound, which is enough to produce an
explosion without transition to detonation. Generally, 3D-models
are found to be more energetic. Multi-dimensional calculations
of nucleosynthesis in the pure deflagration Chandrasekhar model
\cite{Travaglio&al2004} shows that turbulent flame converts
about $50\%$ of carbon and oxygen to ash with different
composition depending on the density of the unburned material.
To burn most of the material in the center (so that to avoid
unobserved low-velocity carbon, oxygen and intermediate-mass
elements in the spectra), the model requires a large number of
ignition spots. In contrast, 3D simulations of the
delayed-detonation SN Ia explosion \cite{Gamezo&al2004} indicate
that there is no such problem in this model and the explosion
energy is higher than that obtained in the pure deflagration
burning. It remains to be seen from future observations which
model is more adequate.

Note that uncertainties in the ignition conditions of the
degenerate star leads to some irreducible diversity of the
explosion kinetic energy, peak luminosity, nickel production for
the same initial configuration. Modeling of light curves of SN
Ia turns out to be a powerful tool to check the SN Ia explosion
models (see recent calculations by multi-group radiation
hydrocode STELLA \cite{Blinnikov&Sorokina2004}).

\section{Gamma-ray bursts}

GRBs have remained in the focus of modern astrophysical 
studies for more than 30 years. After the discovery of GRB afterglows
in 1997 \cite{Costa&al1997}, the model of GRB as being due to 
a strong explosion with isotropic energy release of $10^{53}$ 
ergs in the
interstellar medium 
became widely recognized. 
Various aspects of GRB phenomenology 
are discussed in many reviews: observational and theoretical 
studies are summarized in \cite{Hurley&al2003}, 
first observations of afterglows are
specially reviewed in \cite{jvP&al2000}, GRB theory 
is extensively discussed in \cite{Meszaros2002, Zhang&Meszaros2004}.  

A widely used paradigm for GRBs is the so-called fireball model
(e.g. reviews by Piran \cite{Piran2000, Piran2004} and
references therein). In this model, the energy is released in
the form of thermal energy (its initial form is usually not
specified) near the compact central source (at distances and is
mostly converted into leptons and photons (the fireball itself).
The relativistic outflow (wind) is formed driven by the high
photon-lepton pressure (generically in the form of two
oppositely directed narrow collimated jets) \cite{Paczynski1990,
Shemi&Piran1990}. The fireball internal energy is converted to
the bulk motion of ions so that relativistic speed with high
Lorentz-factors (typically, $\Gamma > 100$) is achieved during
the initial stage of the expansion; the ultrarelativistic motion
is in fact dictated by the need to solve the fireball
compactness problem (see \cite{Blinnikov2000} for a detailed
discussion and references). The kinetic motion of ions is
reconverted back into heat in strong collisionless relativistic
shocks at typical distances of $10^{12}$ cm. Assuming the
appropriate turbulence magnetic field generation and particle
acceleration in the shocks, energy thermalized in the shocks is
emitted via synchrotron and inverse-Compton radiation 
of shock-accelerated electrons
\cite{Rees&Meszaros1992} (see \cite{Waxman2003} for a review),
which is identified with the GRB emission. A shell of
ultrarelativistically moving cold protons produces a blast wave
in the surrounding medium, forming an external shock propagating
outward and reverse shock that propagates inward and decelerates 
the explosion debris. Most energy of the explosion is now carried by the
external shock which decelerates in the surrounding medium.
Assuming magnetic field generation and particle acceleration in
the external shock, the afterglow synchrotron emission of GRB is
produced in radio
\cite{Paczynski&Rhoads1993}, optical \cite{Katz1994, Meszaros&Rees1997}
and X-rays \cite{Vietri1997, Meszaros&Rees1997}. Note that at
this stage the memory of the initial explosion conditions is
cleaned, and the dynamical evolution of the external shock is
well described by the Blandford-McKee self-similar solution
\cite{BMK1976}, a relativistic analog of the
Sedov-von-Neumann-Taylor solution for strong point-like
explosion. This explains the apparent success in modeling the
GRB afterglow spectral and temporal behavior in the framework of
the simple synchrotron model \cite{Wijers&al1997}, irrespective
of the actual nature of the GRB explosion.

There is no consensus thus far about the origin of
the GRB emission itself. Within the fireball model, the GRB 
can be produced by internal fireball dissipation (the internal shock wave model, 
e.g. \cite{Rees&Meszaros1994}), or in the external blast wave decelerating
in the ambient (inhomogeneous) medium 
\cite{Meszaros&Rees1993, Dermer&Mitman1999}. 
The fireball model is known to face some important 
problems (for example, baryon contamination of the fireball, the
microphysics of magnetic field generation and particle acceleration 
in collisionless ultrarelativistic shocks etc., see a 
critical review in \cite{Lyutikov&Blandford2003, Dermer2004}). 
In Ref. \cite{Lyutikov&Blandford2003} an 
alternative to the fireball model is analyzed in which large-scale 
magnetic fields are dynamically important. Whether the GRB jets are hot
(fireball model) or cold (electromagnetic model) remains to be determined
from future observations. Here crucial may be spotting the very early 
GRB afterglows and measuring polarization of prompt GRB emission (see
\cite{Lyutikov2004} 
for the short-list of the electromagnetic model predictions). Note that
irrespective of the mechanism mediating the 
energy transfer from the central source to the
baryon-free region, many essential features of the observed non-thermal GRB
spectra can be reproduced in some general physical models of
prompt gamma-ray emission of GRBs, e.g. by synchrotron
self-Compton emission of plasma with continuously heated nearly monoenergetic
electrons \cite{Stern&Poutanen2004}.

An important open issue is whether GRBs can be the sources 
of ultra-high energy cosmic rays (UHECR). This association 
was suggested in Refs. \cite{Waxman1995, Milgrom&Usov1995, Vietri1995}
based on similarity of the energy
release in GRBs $\sim 10^{44}$ erg per year per cubic Megaparsec 
with what is observed in UHECRs and assuming 
effective proton acceleration in relativistic collisionless shocks. 
The mechanism of UHECR production in GRBs is still uncertain, but 
basic requirements for proton acceleration to high energies
in mildly-relativistic
shocks (pertinent to the internal shock model of GRBs) appear to be  
satisfied. See Ref. \cite{Waxman2004} for more detail and 
discussion. See also lectures by M. Teshima and M. Ostrowski on 
this School.  

Below we focus on the observed association of 
GRBs with an energetic subclass of core-collapse supernovae, 
type Ibc SNe, which with each new finding provides an increasing
evidence that the GRB phenomenon is related to the evolution of
most massive stars and formation of relativistic compact objects
(neutron stars and black holes). 

\section{Supernova - GRB connection} 

\subsection{Theoretical grounds: the collapsar model}

The connection of GRBs with stellar explosions was first proposed
theoretically. Woosley (1993) \cite{Woosley1993} considered a model 
of accretion onto a newly formed rotating black hole to power
the GRB fireball. The progenitor to GRB in this model is a rapidly rotating
Wolf-Rayet (WR) star deprived of its hydrogen and even helium envelop 
due to powerful stellar wind or mass transfer in a binary system. 
Dubbed by Woosley himself as "failed type Ib supernovae", this model is now 
called the collapsar model \cite{MacFadyen&Woosley1999}. 
In this model, a  massive ($\gtrsim 25 M_\odot$) rotating star 
with a helium core $\gtrsim 10 M_\odot$  
collapses to form a rapidly rotating BH with 
mass $\gtrsim 2-3 M_\odot$. The accretion disk from the 
presupernova debris around the BH is assumed to be the energy 
source for GRB and is shown to be capable of providing the prerequisite 
$10^{51}-10^{52}$ ergs via viscous dissipation into neutrino-antineutrino
fireball. The energy released is
assumed to be canalized in two thin antiparallel jets penetrating
the stellar envelope. 

Another possible energy source in the collapsar 
model could be the electromagnetic (Poynting-dominated) beamed 
outflow created via MHD processes, much alike what happens 
in the active galactic nuclei powered by accretion onto a supermassive
BH. The estimates show that  
the Blandford-Znajek (BZ) process \cite{BZ1977} in the collapsar model
(e.g. \cite{Lee&al2000}) can be a viable candidate for 
the central engine mechanism for 
GRBs, provided somewhat extreme values for BH spin (the Kerr parameter
$a=Jc/GM^2\sim 1$) and magnetic field strength in the inner accretion 
disk around the BH ($B\sim 10^{14}-10^{15}$ G). In that case the 
rotating energy of BH (up to $0.29 M_{bh}c^2$ for $a=1$) 
is transformed to the Poynting-dominated jet with energy sufficient to 
subsequently produce a GRB.  

Yet another source of energy in the collapsar model could be the rotation
energy of a rapidly spinning neutron star with high magnetic field
(magnetar), as originally proposed by Usov \cite{Usov1992, Usov1994}. As in
the BZ-based models, the GRB jets are Poynting-dominated. Lyutikov and
Blandford \cite{Lyutikov&Blandford2003} develop the electromagnetic model,
which postulates that the rotating energy of the GRB central engine is
transformed into the electromagnetic energy (for example, in a way similar
to the Goldreich-Julian pulsar model) and is stored in a thin
electromagnetically-dominated "bubble" inside the star. The bubble expands
most rapidly along the rotational axis, breaks out of the stellar envelopes
and drives the ultrarelativistic shock in the circumstellar material. In
contrast to the synchrotron GRB model, here GRB is produced directly by the
magnetic field dissipation due to current-driven instabilities in this shell
after the breakout. The energy transfer to GRB is mediated all the way by
electromagnetic field and not by the ion bulk kinetic energy. It remains to
be checked by observations whether the EM or fireball model for GRB emission
is correct.

A different scenario is the so-called "supranova" model for GRB proposed in
\cite{Vietri&Stella1998} involves a delayed collapse to BH through
the formation of a super-massive rotationally supported neutron star (see
also \cite{Vietri&Stella1999} for another variant of this model). In the
supranova scenario GRB is associated with the BH formation and occurs in
baryon-clean surrounding after the initial supernova explosion. The original
scenario predicts several weeks - month time delay between the SN explosion
and GRB. The critical comparison of the supranova mechanism with
observations (showing its inconsistency) can be found in \cite{Dermer2004}.
In this paper an interesting 
extension of the supranova scenario is considered in which the
second collapse occurs minutes or hours after the primary SN explosion. In
this model, a rotationally-supported magnetized neutron star is formed in a
SN type Ic explosion and rapidly loses energy via powerful pulsar wind along
the rotational axis. The wind drills the baryon-clean polar cones through
which a newly formed rapidly spinning BH generates relativistic jets
producing a GRB (for example, by BZ mechanism). 
Though no rapid rotation of magnetized neutron star  
has been obtained in stellar evolution calculations 
(nor can it definitely be inferred from the existing observations, cf.
\cite{Janka&al2004b}), the 
spin increase with the remnant's mass found in Ref.
\cite{Heger&al2004} may provide credence to this mechanism.  
Some signatures from the
pulsar-wind heated supernova shell can be tested in future observations. 

\subsection{Observational evidence: GRB-supernova associations} 
\label{sec:sn-grb}
First hint on the association of GRBs with SNe came from 
the apparent time coincidence (to within about a day) 
of GRB 980425 with a  
peculiar supernova SN 1998bw \cite{Galama&al1998}.
SN 1998bw occurred in a spiral arm of  
nearby (redshift $z=0.0085$, distance $\sim 40$ Mpc) 
spiral galaxy ESO 184-
G82. 
Such a close location of GRB 980425 rendered it a 
significant outliers by (isotropic) energy release 
$\Delta E_{iso}\approx 10^{48}$ erg from the bulk of 
other GRBs with known energy release, and even from 
a beaming-corrected mean value of GRB energies of $\sim 10^{51}$ erg
\cite{Frail&al2001}.  

Now the most convincing evidence for GRB-SN association 
is provided by spectroscopic observations of expanding
photosphere features in late 
GRB afterglows. Especially strong is the 
case of a bright GRB 030329 associated with SN 2003dh
\cite{Hjorth&al2003, Stanek&al2003, Matheson&al2003, Mazzali&al2003, 
Kawabata&al2003}. Spectral observations of the optical
afterglow of this GRB revealed the presence of
thermal excess above non-thermal power-law continuum  typical
for GRB afterglows. Broad absorption troughs which became more
and more pronounced as the afterglow faded indicated the presence of
high-velocity ejecta similar to those found in 
spectra of SN 1998bw. Despite these strong evidences, 
there are some facts which cannot be explained by simple
combination of the typical SN Ibc spectrum and non-thermal power-law continuum.
For example, the earliest spectroscopic observations of GRB 030329 of 
optical spectra taken on the 6-m telescope SAO RAS 
10-12 hours after the burst \cite{Sokolov&al2004} showed the presence
of broad spectral features which could not be produced by a SN 
at such an early stage. The complicated shape of the 
optical light curve of this GRB with many rebrightenings 
\cite{Lipkin&al2004} and 
polarization observations made by VLT \cite{Greiner&al2003} suggest 
a clumpy circumburst medium and require additional refreshening of
shocks (if one applies the synchrotron model, e.g. \cite{Granot&al2003}).   

Another interesting 
example of GRB-SN connection is provided by GRB
031203. This GRB is one of the closest ($z=0.105$) known GRBs and is found
to be intrinsically faint, $\Delta E_{iso}\sim 10^{50}$ ergs 
(\cite{Watson&al2004, Sazonov&al2004}
\footnote{A bright soft X-ray flux was inferred from XMM observations
of evolving X-ray halo for this burst \cite{Vaughan&al2004}, making it 
an X-ray rich GRB \cite{Watson&al2004}; 
this point of view was questioned by \cite{Sazonov&al2004}.}. 
The low energy release in gamma-rays is confirmed by the 
afterglow calorimetry derived
from the follow-up radio observations \cite{Soderberg&al2004} and
allows this GRB to be considered as an analog to GRB 980425. It is important 
that the low energy release in these bursts can not be ascribed to the 
off-axis observations of a "standard" GRB jet (unless one assumes 
a special broken power-law shape of GRB luminosity function, 
see \cite{Guetta&al2004}). However, a
bright type Ib/c supernova SN 2003lw was associated with GRB 031203 as
suggested by the rebrightening of the R light curve peaking 18 days after
the burst and broad features in the optical spectra taken close to the
maximum of the rebrightening \cite{Cobb&al2004, Thomsen&al2004, 
Malesani&al2004, GalYam&al2004}.

The comparison of radio properties of 33 SNe type Ib/c with those 
of measured radio GRB afterglows allowed Berger et al. \cite{Berger&al2003} 
to conclude that not more than few per cents of SNe type Ib/c could be 
associated with GRBs, which explains the observed small 
galactic rate of GRBs. However, it still remains to be studied how much 
intrinsically faint GRBs like 980425 and 031203 can contribute to the
total GRB rate. 

\section {Hypernovae}

Core-collapse supernovae with kinetic energy of the ejecta 
$\sim 10-30$ times as high as the standard 1 foe ($1\hbox{foe}=10^{51}$ erg)
are now collectively called "hypernovae". The term was introduced by 
B. Paczynski shortly after the discovery of first GRB afterglows in 1997 
by the Beppo-SAX satellite \cite{Paczynski1998} based on qualitative 
analysis of possible evolutionary ways leading to cosmic GRB explosions. 

SN 1998bw was exceptionally bright compared to other Ib/c SNe (the peak
bolometric luminosity of order $10^{43}$ erg/s, comparable to the SN Ia peak
luminosities). This points to the presence of a substantial amount of
$^{56}$Ni isotope, the radioactive decay thereof being thought to power the
early SN light curves. The spectra and light curve of SN 1998bw was modeled
by the explosion of a bare C+O of a very massive star that has lost its
hydrogen and helium envelopes with a kinetic energy more than ten times
typical SNe energies\cite{Iwamoto&al1998}, and they called SN 1998bw a
hypernova.

Since then several other SNe were classified as SN 1998bw-like hypernovae by
their spectral features and light curves: SN 1997ef, SN 2002ap, SN
2003dh/GRB030329, SN 2003lw/031203. Recently, SN 1997dq was dubbed a
hypernova by its similarity with SN 1997ef \cite{Mazzali&al2004}.
   
Extensive numerical modeling of light curves and spectra of hypernovae (see
\cite{Nomoto&al2004} for a recent review) confirmed the need of atypically
high for core-collapse SNe mass of nickel ($\sim 0.1-0.5 M_\odot$) to be
present in the ejecta in order to explain the observed hypernova properties.
The rapid rise in of the observed light curves of the "canonical" SN 1998bw
requires a substantial amount of $^{56}$Ni to be present near the surface.
This strongly indicates the important role of mixing during the explosion as
nickel is synthesized in deep layers during a spherical explosion. This fact
can serve as an additional evidence for non-spherical type Ic explosions. As
we already stressed above, the asphericity appears to be a ubiquitous
feature of core-collapse supernovae in general, culminating in 
bipolar hypernova explosions associated with GRBs.

Spectral modeling suggests \cite{Nomoto&al2004} that the broad-band spectral
features generally seen in early and maximum light of hypernovae signal very
rapid photospheric expansion. In particular, authors of Ref.
\cite{Nomoto&al2004} notice the very unusual for
other SNe fact that OI ($\lambda =7774 A$) and CaII IR (at $\lambda \sim
8000 A$) absorption lines merge into a single broad absorption in early
spectra of SN 1998bw, which indicates a very large velocity of the ejecta
(the line separation $\sim 30000$ km/s).

In general, varying (a) the progenitor C+O core mass from 2 to $\sim 14$
solar masses, choosing (2) the appropriate mass cut (corresponding to the
mass of the compact remnant, a neutron star or black hole $M_c=1.2-4
M_\odot$), and (3) mass of $^{56}$Ni isotope ($\sim 0.1-0.5 M_\odot$) and
its mixing allow \cite{Nomoto&al2004} to reproduce the observed spectra and
light curves of hypernovae.

The analysis of nucleosynthesis in hypernovae suggests a possible
classification scheme of supernova explosions \cite{Nomoto&al2003}. 
In this scheme, core collapse
in stars with initial main sequence masses 
$M_{ms}<25-30 M_\odot$ leads to the formation of neutron stars, while more
massive stars end up with the formation of black holes. Whether or not the
collapse of such massive stars is associated with powerful hypernovae
("Hypernova branch") or faint supernovae ("Faint SN branch") can depend on
additional ("hidden") physical parameters, such as the presupernova 
rotation, magnetic fields.
\cite{Ergma&vdH1998}, or the GRB progenitor being a 
massive binary system component \cite{Tutukov&Cherepashchuk2003,
Podsiadlowski&al2004}. The
need for other parameters determining the outcome of the core collapse also
follows from the continuous distribution of C+O cores of massive
stars before the collapse, as inferred from observations, 
and strong discontinuity between masses of
compact remnants (the "mass gap" between neutron stars and black holes)
\cite{Cherepashchuk2001}\footnote{Recent radial velocity measurements of 
the companion star in low-mass X-ray binary 2S 0921-630 limits the 
mass of the compact object within the range $1.9\pm 0.25 M_\odot< M_x<
2.9\pm 0.4 M_\odot$, making it the plausible high-mass neutron star or 
low-mass BH \protect\cite{Jonker&al2004}. A slightly higher mass 
interval $2.0M_\odot < M_x < 4.3 M_\odot$ (1-sigma) was obtained in 
\protect\cite{Shahbaz&al2004}.}. 

The mass of $^{56}$Ni synthesized in core collapse also appears to correlate
with $M_{ms}$. In ordinary SNe (like 1987a, 1993j, 1994i), $M_{Ni}=0.08\pm
0.03 M_\odot$, but for hypernovae this mass increases up to $\sim 0.5
M_\odot$ for the most energetic events. Large amount of $^{56}$Ni in
hypernovae suggests a different nucleosynthesis event. It was shown that
unlike conventional core-collapse shock nucleosynthesis
\cite{Woosley&Weaver1995}, nucleosynthesis in bipolar supernova explosions
\cite{Maeda&Nomoto2003} or in relativistic modest-entropy massive
wind from accretion disk around a BH \cite{Pruet&al2004} can in 
principle give the observed high amount of nickel in hypernovae. 
Another important consequence of hypernova nucleosynthesis 
can be larger 
abundances (relative to the solar one) 
of Zn, Co, V and smaller abundance of Mn, Cr, the enhanced ratios of
$\alpha$-elements, and large ratio of Si, S relative  to oxygen
(see \cite{Nomoto&al2003} for further detail). 

It is also necessary to note that energy requirements for 
hypernova explosions (
$\Delta E>2\times 10^{51}$ ergs) can hardly be provided by the most elaborated
delayed neutrino explosion mechanism. Indeed, the net explosion energy 
in this mechanism comes mostly from nuclear recombination of matter
inside the gain radius. Analytical \cite{Janka2001} and numerical 
calculations \cite{Janka&al2003} suggest this mass to fall within 
$0.01-0.1 M_\odot$ range, so assuming the recombination 
energy release 8 MeV per nucleon results in $\sim 10^{51}$ ergs of
the explosion energy (see also 
calculations of the SN explosion energy as a function of the 
progenitor mass in Ref. \cite{Fryer1999}). 
The magnetorotational mechanism \cite{Ardeljan&al2004b}
fails to produce the hypernova energies either (see above). Unless 
some non-standard physics does operate, the hypernova energies can be 
recovered from accretion of the rotating collapse debris 
onto BH ($\Delta E\sim 0.06-0.42 \Delta M_a c^2$ 
depending on the BH spin) or by BZ mechanism ($\Delta E \leq 0.29
M_{bh}c^2$). In both cases, rapid rotation (and extremely large magnetic
field for BZ to operate) of the presupernova is 
required. Here realistic self-consistent calculations still 
have to be done, which is a very difficult task. 

\section{Progenitors of GRBs}

The GRB-SN connection leads to the generally accepted 
concept that massive stars that lost their envelopes are progenitors of
long GRBs (this limitation is due to the fact that predominantly long GRBs
with duration > 2 s can be well localized on the sky and provide
rapid alerts for follow-up multiwavelength observations). For short 
single-pulsed GRBs (a quarter of all bursts, see e.g. catalog by Stern et al. 
\cite{Stern&al2001}) the binary NS+NS/NS+BH merging hypothesis 
\cite{Blinnikov&al1984, Ruffert&Janka1999, Janka&al1999} 
remains viable (see also recent general relativistic hydrodynamic 
models of the launch and propagation of relativistic 
jets due to thermal energy deposition near the center of binary mergers
\cite{Aloy&al2004}). 

As we already noted, the emerging empirical evidence is that there exist
intrinsically faint, single-pulsed, apparently spherically-symmetric GRBs
(980425, 031203) associated with strong hypernovae. These hypernovae require
maximal amount of nickel to be synthesized in explosion and large kinetic
energies. On the other hand, another unequivocal hypernova SN 2003dh,
associated with the "classical" GRB 030329, can be modeled with
exceptionally high kinetic energy ($4\times 10^{52}$ ergs) but smaller
amount of nickel ($\sim 0.35 M_\odot$) and smaller mass of the ejecta ($8-10
M_\odot$) \cite{Mazzali&al2003}. These parameters were obtained assuming
spherical symmetry, which is of course not the case for GRB
030329\footnote{For example, two-component modeling of SN 2003dh as a slowly
moving high-mass equatorial ejection and almost discontinuous low-mass polar
outflow
\cite{Woosley&Heger2004} requires the high mass of synthesized 
Ni-56 $\sim 0.5 M_\odot$, as in other hypernovae}. 
But if this tendency is real and will be confirmed
by later observations, we can return to our hypothesis 
\cite{Postnov&Cherepashchuk2001} that there 
should be distinct classes of GRBs according to what is the 
final outcome of collapse of the CO-core of a massive star. 
If the collapse ends up with the formation of a neutron star, 
an intrinsically faint smooth GRB could be produced 
and a heavy envelope is ejected in 
the associated SN Ib/c explosion. The GRB energy in this case 
can be essentially the rotation energy of the neutron star $\sim 10^{49}-10^{50}$ ergs, 
as in the electromagnetic model \cite{Usov1992}.  
If a BH is formed, a lighter envelope is ejected 
with accordingly smaller amount of nickel and possibly 
with higher kinetic energy of the ejecta,  
and more energetic, highly variable 
GRB with a "universal" jet structure \cite{Postnov&al2001} emerges
fed by non-stationary accretion onto the BH. 

The GRB energy can be also interpreted in more exotic way
requiring a new physics. The possible relation of GRBs to mirror
dark matter was discussed in Ref.
\cite{Blinnikov2000}. The conversion of light axions from SNe to
photons as the source of the GRB fireballs was considered in
\cite{Loeb1993, Bertolami1999}.
Recently it was suggested 
\cite{Gianfanga&al2004} that 
ultramassive axions in the mirror world with the Peccei-Quinn
scale $f_a\sim 10^4-10^6$ GeV and mass $m_a\sim 1$ MeV can be
produced in the gravitational collapse or in the merging of two
compact stars. The axions tap most of the released energy and
can decay $\sim 1000$ km away mostly into visible
electron-positron pairs (with $100\%$ conversion efficiency)
thus creating the primary GRB fireball. 
The estimates show that
successful short GRBs can be obtained in compact binary
coalescences, while long GRBs can be created in collapsars. In
extended SN II progenitors, this energy may help the mantle
ejection. In compact CO-progenitors for SN Ib/c axions decay
inside the star, so depending on the stellar radius weaker or
stronger GRBs associated with SNe type Ib/c explosions can be
observed. In this picture again the collapse with the formation
of a neutron star or BH may have different signatures.
   
\section{Conclusions}

Cosmic explosions including various types of supernovae and GRBs
are natural consequence of stellar evolution. The big efforts of
different groups to theoretically understand the physical
mechanism(s) of core collapse SNe appear to be approaching the
final phase. The thermal (neutrino-driven) mechanism for core
collapse SNe is mostly elaborated at present but still fails to
produce a strong explosion. There is understanding why stars do
not explode by this mechanism, and the road map is designed how
to obtain the explosion. Recent multidimensional simulations
with accurate incorporation of Boltzmann neutrino transport
indicate that effectively a modest boost in the neutrino
luminosity is still required. This can be due to inclusion of
new fluid instabilities and more accurate treatment of neutrino
processes and microphysics (the neutron star equation of state)
into the full 3D hydrodynamic calculations. The rotation and
magnetic fields is not yet fully included into calculations,
which is a challenging task. First results of 2D MHD
calculations of the magnetorotational supernova explosion seem
to be encouraging. The most energetic (hypernova) explosions,
however, require an additional to neutrino source of energy, and
the rotation and magnetic fields can be the principal
ingredients.

An impressive progress has been done in multidimensional calculations of
thermonuclear explosions of degenerate dwarfs for type Ia supernovae. It is
still however unclear whether pure deflagration or delayed detonation is at
work in SN Ia. The important problem is to more precisely determine the
initial ignition conditions. Detailed radiation hydrodynamic modeling
revealed that SN Ia light curves proved to be very sensitive to the
explosion models and thus can be used to check the models.
  
In contrast to SNe, the nature of cosmic GRBs remains unclear.
The most important recent progress in understanding GRBs was in 
establishing the link of at least part of them to unusually
energetic SN Ibc explosions (hypernovae). At present several
unequivocal GRB-SN associations are known. The two closest GRBs
discovered so far (GRB 989425 and GRB 031203) proved to be intrinsically
weak compared to the bulk of other GRBs with measured redshifts.
They both show a single-peak smooth gamma-ray light curve with
no signs of jet-induced breaks in the afterglows. In the third
(most strong) case of the GRB-SN association, GRB 030329/SN
2003dh, the GRB light curve is two-peaked, the afterglows show
evidence for jet. Modeling of the underlaid hypernovae light
curve and spectra
 revealed the first two cases to require smaller kinetic
energies but higher mass of the ejecta and the amount of the
synthesized nickel than SN 2003dh. We tentatively propose that
the tendency "weaker, more spherically symmetric GRB - stronger
hypernova" may indicate the formation of a NS in the case of
weak GRBs and of a BH in the case of strong variable GRBs as the
final outcome of the core collapse. In the NS case the GRB
energy comes from the rotational energy of neutron star and is
possibly mediated by the electromagnetic field. When BH is
formed the GRB energy source is the gravitational energy
released during non-stationary accretion onto the black hole or
the black hole rotation. It is not still excluded that the GRB
phenomenon signals some new physics underlying the formation of
compact stars.
 
Observations of various types of supernovae in other galaxies 
and especially 
of a (long awaited!) galactic event by all available means
(including neutrino and gravitational wave detectors) 
should undoubtedly 
be crucial for further understanding physical mechanisms of
cosmic explosions. 
We are sure that the increasing statistics of GRB/SNe in the
nearest future obtained with new GRB-dedicated space missions
like SWIFT will tell us much more on the nature of GRBs and
their progenitors.

\begin{acknowledgments}
The author acknowledges the support through
RFBR grants 02-02-16500, 03-02-17174 and 04-02-16720.
\end{acknowledgments}

\end{document}